# Beyond gold nanoparticles cytotoxicity: potential to impair metastasis hallmarks

Jenifer Pendiuk Gonçalves[1], Anderson Fraga da Cruz[1], Heloise Ribeiro de Barros[2,3], Beatriz Santana Borges[4], Lia Carolina Almeida Soares de Medeiros[4], Maurilio José Soares[4], Mayara Padovan dos Santos[5], Marco Tadeu Grassi[5], Anil Chandra[6], Loretta Laureana del Mercato[6], Gustavo Rodrigues Rossi[1], Edvaldo da Silva Trindade[1], Izabel Cristina Riegel Vidotti[2], Carolina Camargo de Oliveira[1]*.

[1] Laboratory of Inflammatory and Neoplastic Cells/ Laboratory of Sulfated Polysaccharides Investigation, Cell Biology Department, Section of Biological Sciences - Universidade Federal do Paraná, Brazil - Av Cel Francisco H dos Santos, s/n – CEP 81530-980 – Curitiba/PR. Phone: +55 (41) 3361-1770.

[2] Macromolecules and Interfaces Research Group, Department of Chemistry - Universidade Federal do Paraná, Brazil - Av Cel Francisco H dos Santos, s/n – CEP 81530-980 – Curitiba/PR. Phone: +55 (41) 3361-3184.

[3] Laboratory of Electroactive Materials, Chemistry Institute – Universidade de São Paulo – Av Professor Lineu Prestes, 748 – CEP 05513-970 São Paulo/SP.

[4] Laboratory of Cell Biology - Instituto Carlos Chagas (ICC/Fiocruz), Brazil - Rua Prof. Algacyr Munhoz Mader, 3775 - CEP 81350-010 - Curitiba/PR. Phone: +55 (41) 3316-3230.

[5] Environmental Chemistry Group, Department of Chemistry - Universidade Federal do Paraná, Brazil - Av Cel Francisco H dos Santos, s/n – CEP 81530-980 – Curitiba/PR.

[6] CNR NANOTEC – Institute of Nanotechnology c/o Campus Ecotekne, via Monteroni, 73100, Lecce, Italy Phone: + 39 0832319825.

* Corresponding author: Section of Biological Sciences, Cell Biology Department, room 195-B - Universidade Federal do Paraná, Brazil - Av Cel Francisco H dos Santos, s/n – CEP 81530-980 – Curitiba/PR. Phone: +55 (41) 3361-1770. E-mail: krokoli@ufpr.br

**ABSTRACT**

Gold nanoparticle (AuNP)-based systems have been extensively investigated as diagnostic and therapeutic agents due to their tunable properties and easy surface functionalization. Upon cell uptake, AuNPs present an inherent cell impairment potential based on organelle and macromolecules damage, leading to cell death. Such cytotoxicity is concentration-dependent and completely undesirable, especially if unspecific. However, under non-cytotoxic concentrations, internalized AuNPs could potentially weaken cells and act as antitumor agents. Therefore, this study aimed to investigate the antitumor effect of ultrasmall AuNPs (~3 nm) stabilized by the anionic polysaccharide gum arabic (GA-AuNPs). Other than intrinsic cytotoxicity, the focus was downregulation of cancer hallmarks of aggressive tumors, using a highly metastatic model of melanoma. We first demonstrated that GA-AuNPs showed excellent stability under biological environment. Non-cytotoxic concentrations to seven different cell lines, including tumorigenic and non-tumorigenic cells, were determined by standard 2D *in vitro* assays. Gold concentrations $\leq$ 2.4 mg L$^{-1}$ (16.5 nM AuNPs) were non-cytotoxic and therefore chosen for further analyses. Cells exposed to GA-AuNPs were uptaken by melanoma cells through endocytic processes. Next we described remarkable biological properties using non-cytotoxic concentrations of this nanomaterial. Invasion through an extracellular matrix barrier as well as 3D growth capacity (anchorage-independent colony formation and spheroids growth) were negatively affected by 2.4 mg $L^{-1}$ GA-AuNPs. Additionally, exposed spheroids showed morphological changes, suggesting that GA-AuNPs could penetrate into the preformed tumor and affect its integrity. All together these results demonstrate that side effects, such as cytotoxicity, can be avoided by choosing the right concentration, nevertheless, preserving desirable effects such as modulation of key tumor cell malignancy features.

## INTRODUCTION

A major cancer treatment challenge has always been specifically targeting the tumor, thereby reducing side effects of treatment toxicity to healthy tissues as well as tumor resistance to drugs and their availability time in circulation [1]. In this context, advances in tailored treatments allied to nanotechnology are emerging as future therapies [2]. Sophisticated nanotechnology systems engineered to precisely deliver tumor-aimed therapeutic drugs or genes in attempt to avoid undesirable effects [3] are conceptually perfect. Gold nanoparticles (AuNPs) stand out for this purpose. Their extensive surface area, which is easy to functionalize, allows the conjugation of diverse classes of molecules (drugs, genetic silencing tools, ligands for tumor cell receptors, antibodies, etc.) [4,5], making them important tools for the development of controlled drug delivery systems for cancer treatment. Upon light incidence, AuNPs can also be used for photothermic therapy. Due to their surface plasmon resonance, AuNPs can convert light energy into heat, which can be used to destroy cancer cells *in situ* or release therapeutic molecules attached to the nanoparticle´s surface [6–9]. The vast majority of the literature addresses AuNPs in the management and treatment of cancer using one or more of the above shown strategies. Although pre-clinical studies have described great efficacy, and several clinical trials were initiated, the complexity of such developed systems makes them unfeasible for large-scale production and consequently for clinical application [10].

On the other hand, many *in vitro* studies show that an intrinsic antitumor effect can be attributed to AuNPs [11]. Nanoparticles with the ability to conduct electrons (such as AuNPs) have the possibility to interact with active sites of electron donation and reception in cells, such as those present in mitochondria. The capture of electrons by AuNPs can lead to the production of large amounts of reactive oxygen-derived species. Consequently, it affects mitochondrial membrane potential, leading to structural damage to organelle and programmed cell death or even necrosis [12–14].

AuNPs can also damage macromolecules, such as proteins, nucleic acids, membrane lipids, etc. [15]. Such intrinsic antitumor effect of AuNPs is generally strongly associated with the induction of cytotoxicity, which in turn is dependent on the concentration used [16]. However, non-specific cytotoxicity may be dangerous, as it is the strategy behind conventional oncologic chemotherapy and the responsible factor for its severe adverse effects [3,17]. In addition, its effects are often associated with tumor cell resistance development and even secondary neoplasms formation [18–20].

While many studies focus on nanomaterials cytotoxicity and lethality determination [21], little attention is devoted to their therapeutic potential. The many advantages of AuNPs - possibility of manipulating sizes and formats independently and with great precision; relatively simple and well-established aqueous solution synthesis techniques; surface capable of interacting with different classes of molecules and with light in very specific ways [22–24] - place them as important tools for cancer treatment. Our research group have demonstrated that compounds and materials that are non-cytotoxic may still interfere with tumor cells malignant features, weakening them [25,26]. Therefore, this study investigates whether AuNPs may present an intrinsic antitumor effect regardless of cytotoxicity.

The AuNPs system and tumor cell model selected for this study were based on previously published knowledge. Huang and coworkers had shown that ultrasmall AuNPs (<10 nm) exhibit unique advantages over larger ones: more uptake by tumor cells, deeper penetration into multicellular tumor spheroids, and higher intratumoral accumulation *in vivo* [27]. Once inside the cells, ultrasmall AuNPs can reach structures like the nucleus [28] and interfere with DNA transcription [29], which might be important for changing tumor cells malignant features. The conventional citrate-AuNPs systems are electrostatically stabilized by the equally charged ligand molecules on their surface, which can easily be dismissed upon exposure to different salt concentrations or pHs, favoring particles aggregation [2]. For this reason, a colloidal system of ultrasmall

AuNPs stabilized by the anionic polysaccharide gum arabic (GA) was used as a model nanomaterial since this electrostatic and steric stabilizer favors the AuNPs colloidal stability in complex biological environments [30,31]. A previously described and characterized synthesis strategy was used, where AuNPs are simultaneously chemically synthesized and stabilized by GA. The procedure yields on biocompatible ultrasmall (2-5 nm), spherical, well dispersed, and with narrow size distribution AuNPs, that are highly stable as a colloidal suspension in water [30]. Regarding the tumor cell model, our study focused on melanoma since GA-AuNPs exerted a selective influence over metastatic melanoma cells, leading to a decrease in cell viability and proliferation as well as an increase in adhesion of this cell type after a short time exposure in a preliminary study [31]. Therefore, herein we investigated the hypothesis that non-cytotoxic GA-AuNPs concentrations uptaken by melanoma cells *in vitro* could affect functional cellular features that are essential for tumor progression and metastasis. To the best of our knowledge, this is the first study showing the intrinsic therapeutic properties of AuNPs regardless of cytotoxicity against tumor cells.

## METHODS

### Materials

Tetrachloroauric acid (HAuCl$_4$·3H20, 30% in dilute HCl, 99.9%), GA (Mw = 9.3 × 105 g mol$^{-1}$; uronic acid content of 17%), PBS, neutral red dye, MTT, DMSO, rhodamine B isothiocyanate (RBITC), agarose, propidium iodide (PI), saponin and glycine were purchased from Sigma-Aldrich (San Luis, USA). Sodium borohydride (NaBH$_4$, ≥ 98%) and eosin-phloxin were purchased from Nuclear (São Paulo, Brazil). Ultrapure water - 18.2 MΩ cm – (Millipore, Burlington, USA) was used in the preparation of all solutions and in all washing steps. Prior to use, GA powder was solubilized in water, left overnight at 4ºC and subsequently dialyzed for 48 h against distilled water through a dialysis membrane (12–14 kDa cut-off) and freeze-dried.

Formvar was purchased from Ted Pella INC (Redding, USA). Uranyl acetate, glutaraldehyde, cacodylic acid, paraformaldehyde, biobond, osmium tetroxide, hematoxylin and epoxy resin were purchased from Electron Microscopy Sciences (Hatfield, USA). High glucose Dulbecco's Modified Eagle Medium (DMEM) with or without phenol red, fetal bovine serum (FBS), penicillin, streptomycin, trypsin/EDTA, Matrigel, AlgiMatrix™ 3D Culture System and its corresponding firming buffer were purchased from Gibco (Thermo Fisher Scientific, Waltham, USA). All plastic material for cell culture were purchased from Sarstedt (Nümbrecht, Germany); crystal violet dye, EDTA and Entellan from Merk (Darmstadt, Germany); transwell inserts and 10 kDa Amicon-Ultra tubes from Millipore (Burlington, USA); ActinGreen, Lectin WGA 488, Fluoromount-G™ Mounting Medium with or without DAPI, Alexa Fluor 647-conjugated phalloidin and OCT compound from Thermo Fisher Scientific; and DAPI from Invitrogen (Carlsbad, USA).

**GA-AuNPs synthesis and characterization**

AuNPs were chemically synthesized and simultaneously stabilized by gum arabic [30] through the reduction of $HAuCl_4$ ($7.056 \times 10^{-5}$ mol $L^{-1}$) by $NaBH_4$ (0.4 mmol $L^{-1}$) in the presence of GA (0.04% w/v in water). AuNPs localized surface plasmon resonance (LSPR) band was monitored using a Cary 60 UV-vis spectrophotometer (Agilent Technologies, Santa Clara, USA). Surface charge was measured through zeta potential using Zetasizer nano ZS90 (Malvern Panalytical, Malvern, United Kingdom) The obtained suspensions were concentrated (10 times) and washed once with ultrapure water by centrifuging in 10 kDa Amicon-Ultra tubes for 15 minutes at 5000 rpm and 4°C. Gold concentration in the obtained suspensions was determined using an iCAP 6000 series inductively coupled plasma optical emission spectrometer (ICP-OES, Thermo Fisher Scientific, Waltham, USA). Samples were sterilized by filtration against 0.22 μm PVDF membranes and stored at 4°C.

Additional material characterization was performed besides the previously published [30]. For transmission electron microscopy (TEM) characterization, samples were deposited on Formvar-coated copper grids and negatively stained using 2% uranyl acetate [33] in order to determine the spatial interaction between GA and the AuNPs. Air-dried samples were observed using a JEM 1200EX-II (JEOL, Tokyo, Japan) TEM operating at 80 KV. Fiji software [34] was used to measure AuNPs size. From the median AuNPs´ diameter, the volume ($V_{sphere}$) and mass ($M_{sphere}$) of each individual gold nanosphere and the molar concentration of AuNPs in the colloidal suspensions ($N_{mol}$) were respectively calculated using the formulas: $V_{sphere} = 4 \pi r^3 / 3$; $M_{sphere} = V_{sphere} \times D_{Au}$; $N_{mol} = ([Au] / M_{sphere}) / A$, where r = radius (diameter / 2); $D_{Au}$ = gold density ($1.93 \times 10^{-8}$ pg nm$^3$); $[Au]$ = gold concentration in the suspensions measured by ICP-OES; A = Avogadro constant ($6.02 \times 10^{23}$). The nanoparticles´ hydrodynamic size determination in aqueous solution was performed in NanoSight LM10 equipment (Malvern Panalytical, Malvern, United Kingdom) acquiring at 18-28 frames per second at 25ºC.

**GA-AuNPs stability under biological environment**

To simulate the biological environment of an *in vitro* cell culture condition, GA-AuNPs were mixed with complete medium (Dulbecco's Modified Eagle Medium (DMEM) without phenol red, containing 10% fetal bovine serum (FBS)). Mixtures were maintained for 4 days in humid incubator at 37ºC and 5% CO$_2$. After incubation, UV-vis absorption and Small-angle X-ray scattering (SAXS) specters were obtained. Complete medium alone was maintained under the same conditions and used as blank for UV-vis analysis. SAXS measurements were performed on SAXS1 beamline at the Brazilian Synchrotron Light Laboratory to determine nanoparticles size, shape, and polydispersity [30,35]. The scattered X-ray beam with 1.489 Å wavelength (λ) was detected using a Pilatus 300k detector. Sample-to-detector distances were 885.04 and 3015.24 mm to cover a scattering vector, q ($q = (4\,\pi\,/\,\lambda)$ x $\sin\theta$), ranging from 0.03 to 4 $nm^{-1}$, respectively, where $2\theta$ is the scattering angle. Samples normalized scattering image was then subtracted from pure water normalized scattering image, and the isotropic result was radially averaged to obtain I(q) vs. q.

**Cell culture and *in vitro* exposure to GA-AuNPs**

Several tumorigenic and non-tumorigenic cell lines were used: melanoma from mouse (B16-F10 - BCRJ, 0046), and human (CHL-1 - ATCC, CRL-9446™; and A-375 - ATCC, CRL-1619™); melanocytes from mouse (Melan-A - Ximbio, 153599) and human (NGM - BCRJ, 0190); human pancreas carcinoma (MIA PaCa-2 - ATCC, CRL-1420™) and breast cancer (MCF7 - ATCC, HTB-22™). Cells were cultured in each cell´s specific media supplemented with 1 U $mL^{-1}$ penicillin and 1 µg $mL^{-1}$ streptomycin, and 10% FBS. Cells were maintained at 37ºC in humidified atmosphere with 5% $CO_2$ and used for no more than 5 subculturing cycles. Different concentrations of GA-AuNPs were prepared from serial dilutions (1:10 ratio) in 0.4% GA solution (which was also used as vehicle/control for all biological assays). For *in vitro* cell exposure, GA-AuNPs final concentrations ranging from 2.44 µg $L^{-1}$ to 24.4 mg $L^{-1}$ were added to cell culture medium.

**Cell viability assays**

Different cell lines were seeded into 96-well plates at respective densities of cells/$cm^2$ (1.2-1.5 x $10^3$ B16-F10, 4.4 x $10^3$ CHL-1, 1.5 x $10^3$ A-375, 3 x $10^3$ Melan-A, 3 x $10^3$ NGM, 4.5 x $10^3$ MIA PaCa-2 and 4.5 x $10^3$ MCF7). GA-AuNPs were added 6 hours after seeding. Cells grew in the presence of GA-AuNPs or 0.4% GA for 96 hours. Cytotoxicity and cell viability were measured through MTT [36] or neutral red (NR) retention into acidic compartments [37] assays. Briefly, cells were incubated with 0.5 mg $mL^{-1}$ MTT for 3 hours, or with 0.04 mg $mL^{-1}$ NR for 2 hours. The formed formazan crystals were eluted in DMSO and absorbance read at 570 nm. NR retained into cells was eluted in 50% ethanol/1% glacial acetic acid and absorbance read at 540 nm. Cell amount present in each exposed group was estimated by crystal violet (CV) staining [38]. Cells were incubated with 0.25 mg mL$^{-1}$ CV for 20 minutes, dye was eluted in 33% glacial acetic acid, and absorbance read at 570 nm. NR absorbance was normalized by CV absorbance of each sample. Epoch$^{TM}$ Microplate Spectrophotometer (BioTek Instruments, Winooski, USA) was used.

**GA fluorescent labeling (GA-AuNP@RBITC)**

To study cell uptake of GA, the polysaccharide was covalently labeled using RBITC after GA-AuNPs synthesis. Samples were washed in sodium carbonate buffer, pH 8.5, and 0.2 mg mL$^{-1}$ RBITC was added and incubated overnight under constant stirring. The obtained suspension was washed several times with ultrapure water by centrifuging in 10 kDa Amicon-Ultra tubes for 15 minutes at 5000 rpm and 4°C, until all free dye was removed. Final volume was kept equal to initial. Gold concentration in GA-AuNP@RBITC was determined by ICP-OES.

**GA-AuNPs uptake by tumor cells**

For visualization of AuNPs or GA internalization 1.2 x 10$^3$ B16-F10 cells/cm$^2$ were seeded into 175 cm$^2$ culture flasks or 24-well plates over glass coverslips. To detect intracellular AuNPs, cells exposed to 2.44 or 24.4 mg L$^{-1}$ GA-AuNPs for 96 hours were fixed in 2.5% glutaraldehyde, detached using cell scraper, post-fixed in 1% osmium tetroxide, dehydrated with acetone, and impregnated in epoxy resin. Ultra-thin sections were obtained by ultramicrotomy and observed by TEM. To follow GA uptake, coverslip-attached cells were exposed to 2.44 mg L$^{-1}$ GA-AuNP@RBITC for 1 hour at 4 or 37ºC, then washed with DMEM to remove excessive GA-AuNP@RBITC, incubated with 3.3 μg mL$^{-1}$ Lectin WGA 488 for 40 minutes to stain the cell membrane, washed again with DMEM and finally fixed with 2% paraformaldehyde for 20 minutes. Slides were prepared using Fluoromount-G™ Mounting Medium with DAPI. Fluorescence images were taken using A1R MP+ laser scanning confocal microscope (Nikon, Minato, Japan).

For internalized nanoparticles-derived gold quantification, $1.2 \times 10^3$ B16-F10 cells/cm$^2$ were seeded into 75 cm$^2$ culture flasks. After exposure to 2.44 or 24.4 mg L$^{-1}$ GA-AuNPs for 96 hours cells were detached using 2 mM EDTA and fixed in 2% paraformaldehyde for 1 hour. The number of cells in each sample was quantified using a hemocytometer. Cell pellets were resuspended in 250 µL of ultrapure water, mixed with 5 mL of aqua regia (3 HCl : 1 HNO$_3$) and boiled over a heating plate for 10 minutes to digest organic components. Volumes were completed to 25 mL using ultrapure water and dissolved gold concentration measured by ultrasonic nebulizer coupled to ICP-OES (USN-ICP-OES, Agilent 720 ES, Agilent Technologies, Santa Clara, USA). The obtained gold mass from each sample was divided by its respective cell number.

**Tumor cell invasion assay**

Invasive capacity of melanoma cells exposed to GA-AuNPs was determined by Matrigel-barrier invasion assay [39] with modifications. Briefly, 8 µm pores transwell inserts were coated with 35 µL of 2.6 mg mL$^{-1}$ Matrigel, and polymerized under shaking at 30ºC for 1.5 hours. B16-F10 cells suspensions containing $8 \times 10^4$ cells in DMEM were seeded inside each insert in the presence of 0.24, 2.4, or 24.4 mg L$^{-1}$ of GA-AuNPs or 0.4% GA. Complete medium was added to lower chamber as chemoattractant. Cells were incubated for 72 hours, fixed with 2% paraformaldehyde, and labeled with ActinGreen and DAPI. Cells on the membrane top (which did not invade) were mechanically removed. Membranes images were acquired using Metafer VSlide Scanner (MetaSystems, Newton, USA) and Zeiss Axio Imager Z2 microscope (Carl Zeiss, Oberkochen, Germany) and the amount of invasive cells per insert area was determined by nuclei quantification using Fiji software [34].

**Tumor cell 3D growth assays**

Individual melanoma cells capacity to form colonies was evaluated through anchorage-independent growth using AlgiMatrix™ 3D Culture System. Suspensions containing $2.5 \times 10^3$ B16-F10 cells in complete medium were added to the plate in the presence of 0.24, 2.4, or 24.4 mg $L^{-1}$ of GA-AuNPs or 0.4% GA, together with 10% (v/v) firming buffer. Medium was replaced on the third day. Melanoma colonies formed after 6 days were fixed with 1% paraformaldehyde overnight and stained with CV. Excessive dye was washed with distilled water and hydrogels were photographed against checkered-background petri dish, which was further used as scale. Number and size of generated colonies were determined using Fiji software [34]. In parallel, other colonies were harvested from AlgiMatrix by incubation with dissolving buffer, processed for TEM as described before, ultra-thin sections were obtained by ultramicrotomy and observed by TEM to detect the presence of intracellular AuNPs.

To address the effect of GA-AuNPs on preformed 3D structures, melanoma spheroids were produced by low adhesion plate growth. A solution of 2% (w/v) agarose was added to 96-well culture plates (70 μL/well) to cover the cell adherent well bottom. Plates were kept under UV light at room temperature for 30 minutes, then in a cell incubator at 37°C for the same time. Next, $1.6 \times 10^3$ B16-F10 cells/well in complete medium were added on top of the agarose coating. Spheroids were allowed to form during 72 hours. The supernatant was replaced by complete media containing 0.24 or 2.4 mg $L^{-1}$ of GA-AuNPs or 0.4% GA. Medium with GA-AuNPs was replaced every couple of days. After 7 days, spheroids were collected, fixed with 4% paraformaldehyde overnight, and photographed under a light microscope using a hemocytometer chamber as scale. Their size was measured using Fiji software [34].

For histological analysis, spheroids were cryoprotected by consecutive incubations with sucrose solutions for 4 hours each and then embedded in OCT compound, frozen at -80ºC and cryossected (4 µm thickness) using CM1850 equipment (Leica Biosystems, Wetzlar, Germany). Sections attached to biobond-covered histological slides were washed twice with distilled water, stained with hematoxylin for 1 minute, washed again with distilled water, stained with eosin-phloxin for 30 seconds and finally washed twice with distilled water. Materials were dehydrated by incubations in increasing concentrations of ethanol, then xylol. Slides were mounted using Entellan. Images were taken using an AxioLab A1 (Carl Zeiss, Oberkochen, Germany) light microscope. For confocal fluorescence microscopy, spheroids were permeabilized in 0.1% saponin, incubated with 0.1 M glycine, and then labeled with Alexa Fluor 647-conjugated phalloidin and PI. Slides were mounted using FluoromountG and images obtained using A1RMP+ microscope.

**Statistical analysis**

Independent experiments were performed at least three times (except for NanoSight measurement, which was performed once).

GA-AuNPs exposed samples were compared to the respective control (GA 0.4% solution) of each experiment. ROUT and Grubbs tests were applied to identify outliers. Data were submitted to D´Agostino & Pearson and Shapiro-Wilk normality tests, followed by the most appropriate statistic test [40] ($p$ values $< 0.05$ were considered significant).

**RESULTS AND DISCUSSION**

The main objective of this work was to investigate whether GA-AuNPs uptaken by tumor cells may present an intrinsic antitumor effect regardless of cytotoxicity. First, additional GA-AuNPs model characterization and their stability under biological environment were performed in order to better understand their biological effects. Next, standard cell viability assays in a series of tumorigenic and non-tumorigenic cells lines were used to select GA-AuNPs non-cytotoxic concentrations. Microscopy techniques and ICP-OES spectroscopy were used to observe and quantify GA and AuNPs internalization by tumor cells. Finally, the ability of non-cytotoxic GA-AuNPs concentrations to minimize tumor cells key malignant features was verified using 3D *in vitro* assays.

**GA-AuNPs characterization and stability**

GA-AuNPs were synthesized as previously described [30]. Obtained GA-AuNPs showed a LSPR band with maximum absorbance at 512 nm (Supl. Fig. 1A) and median diameter of 2.9 nm (Supl. Fig. 1B). Independent synthesis batches of different volumes (from 30 mL up to 1 L) presented results very similar to each other and to the original description, reinforcing the method´s reproducibility and scalability. The calculated volume and mass of each individual AuNP were 12.8 $nm^3$ and 2.5 x $10^{-7}$ pg, respectively. Zeta potential presented a maximum negative surface charge of -20 mV (Supl. Fig. 1C).

The material model used here consists of ultrasmall AuNPs stabilized by gum arabic. Commercial GA was used, which is commonly extracted from *Acacia Senegal* trees and composed of a mixture of anionic polysaccharides and proteins, with average mass around 1-2 x $10^6$ g $mol^{-1}$ [41]. The study that describes the nanomaterial used here showed that GA acts as a stabilizer by forming nucleation centers for the growth of AuNPs, suggesting a binding between them (as determined by zeta potential measurements) [30]. However, the spatial interaction between both components had not been described. Therefore, to determine GA interaction with AuNPs, negatively stained samples were observed by TEM. An interesting assembled structure is observed: due to the macromolecular structure of GA and its physicochemical characteristics, it appears in the form of colloidal nanometric aggregates [33,42] observed as light rounded particles against a dark background (Fig. 1, A1), which are peripherally surrounded by AuNPs (Fig. 1, A2). Based on obtained TEM images and literature information about GA conformation [42,43], a 3D model for the nanocomposite was suggested (Fig. 1, A3). Considering delivery-efficiency, this carrier pattern can be very interesting to biomedical applications, since most GA particles have several surface-attached AuNPs supplying a set of ultrasmall AuNPs to a biological target, like a tumor cell for example.

Interestingly, TEM images showed that dry GA particles alone or in the presence of AuNPs had different sizes. Therefore, aqueous suspensions of both samples were analyzed and results obtained from NanoSight confirmed that GA particles presented a 16% increase in size when peripherically bounded to AuNPs, but still in nanometric scale (Supl. Fig. 1, D1, and D2). Even AuNPs-unattached GA particles are larger in AuNPs sample compared to GA alone dispersed in water. Therefore, it is possible that changes in the chemical environment created during AuNPs synthesis process led to larger sizes since it could perturb macromolecules hydrogen and hydrophobic bonds as well as electrostatic interactions, being fundamental for their conformation [44].

Previously, GA had been able to maintain AuNPs stability - keeping their morphology, light absorption and aggregation-free profiles - in aqueous solution [30]. Such stability was confirmed in the present study, since AuNPs remain free of sedimentation or precipitation for an indefinite time when stored in refrigerator (data not shown). However, a biological environment condition (temperature, pH, salt concentration, etc.) can influence such stability [44]. Proteins present in biological fluids (such as culture medium containing FBS, blood plasma, among others) can be quickly adsorbed to the nanoparticles surface [45–47]. Aggregation caused in this process can directly affect the interaction of nanomaterials with cells and, consequently, their biological activity. [48–50]. Therefore, to determine GA-AuNPs stability under biological environment, samples were submitted to *in vitro* cell culture conditions for 4 days. First, the UV-visible absorption profile was verified since this is a simple and efficient method of characterizing GA-AuNPs in aqueous solution. No shift in the GA-AuNPs LSPR band position was observed, suffering only a slight enlargement (Fig. 1 C). In addition, SAXS measures were used to correlate GA-AuNPs morphological properties and the effect of their interaction with the culture medium. In similar systems, it has been observed that changes in the scattering profile with low $q$ values are generally attributed to nanoparticles aggregation [51]. Figures 1 D and E show that AuNPs SAXS curves fits (scattering curves and particles diameter distribution) in water and mixed with cell culture medium are quite similar, showing that this particular environment does not promote GA-AuNPs aggregation. The obtained result agrees with the literature, where it was observed that GA-stabilized AuNPs remained stable after 7 days in the presence of biological fluids components, such as histidine, cysteine, albumin and sodium chloride (which are also present in the complete culture medium). The low affinity for plasma proteins and consequent GA-AuNPs stability was attributed to GA´s branched structure of polysaccharides and proteins, which decreases their reactivity, serving as a stable vehicle for AuNPs transport [31].

**GA-AuNPs cytotoxicity and cell uptake**

To test our hypothesis, we have first determined the range of non-cytotoxic GA-AuNPs concentrations in seven different cell lines, tumorigenic and non-tumorigenic, from human and mouse origins. Cells were exposed *in vitro* for 96 hours to concentrations ranging from 2.44 µg L$^{-1}$ to 24.4 mg L$^{-1}$ of gold (as molarity of AuNPs in suspension, the highest concentration used corresponds to 0.165 µM). A 0.4% GA solution was used as vehicle control. Results from MTT assay presented in Fig. 2 showed no cell viability reduction on A-375, NGM, MIA PaCa-2, and MCF7 cells.

However, 24.4 mg L$^{-1}$ reduced B16-F10, CHL-1, and Melan A cell viabilities by 44, 11, and 43%, respectively. Barros and coworkers had shown an apparent specificity of GA-AuNPs to affect melanoma cells [30]. Interestingly, here 24.4 mg L$^{-1}$ GA-AuNPs were cytotoxic only to melanocytic cells (melanoma and melanocyte) and among them, exclusively triple-negative melanomas - which miss targetable driver mutations and consequently are unresponsive to current treatments [52] - were affected. Further detailed studies would be advantageous to understand such cell origin related specificity of AuNPs effects. Also the other less aggressive tumor models might also be responsive to GA-AuNPs and deserve further investigations in the near future.

Since 30% viability loss must be considered toxic [53], and only the highest concentration (24.4 mg L$^{-1}$) was toxic to three out of seven cell lines tested, all the concentration range below that was considered non-cytotoxic. Finally, B16-F10 was selected to continue with the following experiments as the most sensitive metastatic tumor cell line tested.

To guarantee the non-cytotoxicity of the selected concentration range, additional standard viability and proliferation assays were performed on B16-F10 cells upon GA-AuNPs exposure. No statistically significant reduction on cell number was observed (Fig. 3 A) after 96 hours. Results from neutral red uptake assay showed no viability loss but an increase of retained dye (Fig. 3 A and B). These results confirm the selected concentrations non-cytotoxicity and suggest a GA-AuNPs concentration-dependent induction of endocytic processes for their uptake. Consequently, more acidic vesicles are produced inside the cells, causing more neutral red to be retained, since this dye becomes unable to cross cell membranes when protonated [37]. Therefore, the slight reduction on cell number and cytotoxicity at highest tested concentration were probably consequence of GA-AuNPs internalization by cells.

To confirm that tumor cells could internalize GA-AuNPs, exposed melanoma cells were observed by TEM and the uptaken gold mass measured by ICP-OES.

AuNPs are found inside vesicular structures into cells cytoplasm (Fig. 3 C). GA-AuNPs uptake by melanoma cells is concentration-dependent since cells exposed to 24.4 mg $L^{-1}$ present in average 6.8-fold more gold (and as consequence, proportionally more AuNPs) than the ones exposed to 2.44 mg $L^{-1}$ (Table 1). Additionally, fluorescently labeled GA (GA-AuNP@RBITC) are observed inside tumor cells after one hour exposure to 2.44 mg $L^{-1}$. GA-AuNP@RBITC fluorescence shows "granular-like" cytoplasmic distribution and is overlapped to the cell membrane in some regions (Fig. 3 D). Therefore, a non-cytotoxic concentration of GA-AuNPs could be found inside tumor cells, showing that both AuNPs and the stabilizing agent GA – which may also have acted as a carrier - are uptaken. Melanoma cells incubated with GA-AuNPs or GA-AuNP@RBITC at 4ºC show drastically reduced uptake (data not shown), corroborating cell uptake via endocytosis. Based on these previous results, 3 non-toxic concentrations of GA-AuNPs were selected to follow up: 0.024, 0.244, and 2.44 mg $L^{-1}$.

**GA-AuNPs effects on tumor progression-related features of melanoma cells**

The main focus of the present study is to verify whether under non-cytotoxic conditions GA-AuNPs are capable of presenting an intrinsic therapeutic effect on tumor cells malignancy parameters, this is, affecting cellular functions that are essential for tumor progression and metastasis. Metastasis initiates with cell motility/invasion capacity acquisition, evolving to modulate the surrounding microenvironment and being plastic to finally colonize secondary sites [32]. Cell invasion process sets up the first step towards metastasis. Once stablished at a different tissue, cellular ability to homotypically aggregate and grow in colonies reflects their ability to proliferate and ultimately form new tumors [54]. Therefore, melanoma cells were submitted to Matrigel-barrier invasion assay in the presence of GA-AuNPs. Cells exposed to 2.44 mg $L^{-1}$ present 48% reduction in their invasive capacity (Fig. 4 B1 – upper images -, and B2). The acquisition of invasive capacity by tumor cells is the first step towards metastasis, a process that is closely dependent on cell migration. That is, the capacity acquired by cells to leave their original location, move to another region and to degrade the extracellular matrix, allowing them to penetrate the barrier normally conferred by it [55–57]. The obtained images show a clear change in the actin filaments distribution pattern of melanoma cells that show reduced invasive capacity after exposure to 2.44 mg L$^{-1}$ of GA-AuNPs (Fig. 4 B2 – bottom images). The profile of actin accumulation at the edges of cells exposed to GA-AuNPs suggests an increase in their adhesiveness. This result agrees with the previous report that showed melanoma cells more spread out and adhered to the substrate after GA-AuNPs exposure [30].

At the end of the metastatic process, an important characteristic for cancer cells is the ability to aggregate homotypically through glycoproteins and grow in colonies independently of anchoring. This ability reflects their ability to proliferate and finally form new tumors far from origin [54]. Therefore, GA-AuNPs effect on melanoma cells 3D growth was studied. First, the growth of individualized cells into colonies (clonogenic capacity) was evaluated. After 6 days, melanoma cells immersed into 3D matrices in the presence of 2.44 mg L$^{-1}$ GA-AuNPs show 15% less colonies (Fig. 5 B1, and B2). Colonies formed in the presence of 0.24 and 2.44 mg L$^{-1}$ are 22 and 25% smaller than control ones, respectively (Fig. 5 B1, and B3). In addition, colonies formed in the presence of 2.44 mg L$^{-1}$ observed by TEM present GA-AuNPs at the cells surface (Fig. 5 C1), being uptaken by endocytic processes (Fig. 5 C2) and inside the cells (Fig. 5 C3) – mainly surrounded by vesicular structures, but few can be observed free in cytoplasm. There is a good correlation between *in vitro* colony formation assays using tumor cells treated with antitumor agents and the observed *in vivo* results using the same agents [58,59]. Therefore, the results obtained suggest a strong antitumor potential for GA-AuNPs.

Second, melanoma spheroids were exposed to GA-AuNPs as a simulation of preformed tumors. After 7 days, spheroids exposed to 2.44 mg L$^{-1}$ are 12% smaller than controls (Fig. 6 B and C). Histologically, control spheroids present thicker edges, with juxtaposed cells. Their center region is denser in cellular material and present some melanin-producing cells (Fig. 6 D, top line). On the other hand, GA-AuNPs exposed spheroids present thin edges, with loosely connected cells (Fig. 6 D, bottom line) that keep the spheroid integrate while fixed (Fig. 6 E), but disintegrate when the material is sectioned. The center is more porous and contains cells with abnormal morphology (Fig. 6 D, bottom line). Conventional drugs usually diffuse across the extracellular matrix and are easily uptaken by cells. Nanoparticles, however, are more restricted on tissue (as well as tumor and not vascularized 3D models) penetration. They need to be slowly endocytosed and face obstacles when diffusing because of their relative big size [60], but most nanoparticles that penetrate into tumor spheroids are uptaken by cells [61]. In general, ultrasmall (<10 nm) AuNPs are able to penetrate deeply into tumor spheroids [62] and results obtained here show that the central region of exposed spheroids is porous, containing altered cells, characteristic of tumor necrotic areas [63]. Therefore, GA-AuNPs probably penetrated the spheroids, leading to cell distress. Also, the altered proliferative external layer might have accounted for size reduction [64].

Others have previously shown that intracellular vesicles generated to uptake AuNPs disrupt the actin cytoskeleton, affecting the cell contraction process [65]. Results obtained here suggest that this might be a reason for the decreased malignancy cell features, since NR uptake was increased in the presence of GA-AuNPs (indicating endocytic vesicles formation, which has been proved by microscopy images) and actin-associated processes were consequently affected, as cell morphology, migration/invasion, and robustness of cell-cell adhesion – interfering with 3D structures growth.

In summary, we used powerful *in vitro* techniques - that strongly correlate with *in vivo* effects of drugs and nanoparticles on tumor growth and metastasis [59,66] - to determine GA-AuNPs mode of action. GA-AuNPs present intrinsic antitumor potential beyond cytotoxicity. GA-AuNPs were able to reduce both tumor cells capacities to invade and to growth three dimensionally under non-cytotoxic conditions. By weakening tumor cells capacity to complete both initial and final events of metastasis and interfering with preformed 3D tumor structures growth, GA-AuNPs configure as potent candidates to help decelerating the same process *in vivo*. This work broadens the possibility of GA-AuNPs therapeutic application by taking advantage of a simple and economically feasible tool with less - or even no - side effect, using its inherent capacity to impair metastasis-related cellular functions. Therefore, further *in vivo* investigations will be performed to confirm ultrasmall GA-AuNPs safety and intrinsic effectiveness as a cancer nanomedicine. In addition, this study reinforces the thought that there is still a lot to understand about AuNPs mechanisms of action in tumor cells, and the importance to explore possible molecular targets to offer some direction into the changes on transcribed genes or produced proteins. This reveals a new field for systematically exploring the intrinsic antitumor potential of diverse sizes and shapes of AuNPs and other metallic nanoparticles as well.

Authors declare no competing interests.

**AKNOWLEDGEMENTS**

CCdO acknowledges the ERC-CONFAP-CNPq implementing arrangement (127/2018 Fundação Araucária), the UFPR (project 3527/2018 FUNPAR), and UFPR-PROAP (Coordenação de Aperfeiçoamento de Pessoal de Nível Superior - CAPES - Finance Code 001). Authors would like to thank CME-UFPR and CTAF-UFPR for their microscopy services. Dr Roger Chammas for providing A-375, CHL-1, NGM, and MelanA cell lines. Biosciences and Biotechnology Postgraduate Program of ICC/Fiocruz for the NanoSight use. Dr. Éder José dos Santos, Amanda Beatriz

Herrmann and Instituto de Tecnologia do Paraná – Tecpar for USN-ICP-OES measurements. The LNLS for SAXS measurements (project number 20180162) and the SAXS1 beamline staff for the assistance during experiments. Dr. Mateus Borba Cardoso is highly acknowledged for the help with the SAXS fittings. CAPES, CNPq, and UFPR-TN for providing students fellowships.

**FUNDING**

This work was partially supported by the following grants: CAPES-PROAP (PPGBCM-UFPR), CNPq (477467/2010-5), PRPPG-UFPR (04/18 call, project 3527 FUNPAR), the ERC-StG project INTERCELLMED (759959) and the project "TECNOMED - Tecnopolo di Nanotecnologia e Fotonica per la Medicina di Precisione" [Ministry of University and Scientific Research (MIUR) Decreto Direttoriale n. 3449 del 4/12/2017, CUP B83B17000010001].

**DATA AVAILABILITY**

The datasets used and/or analyzed during the current study are available from the corresponding author on reasonable request.

## FIGURES WITH LEGENDS

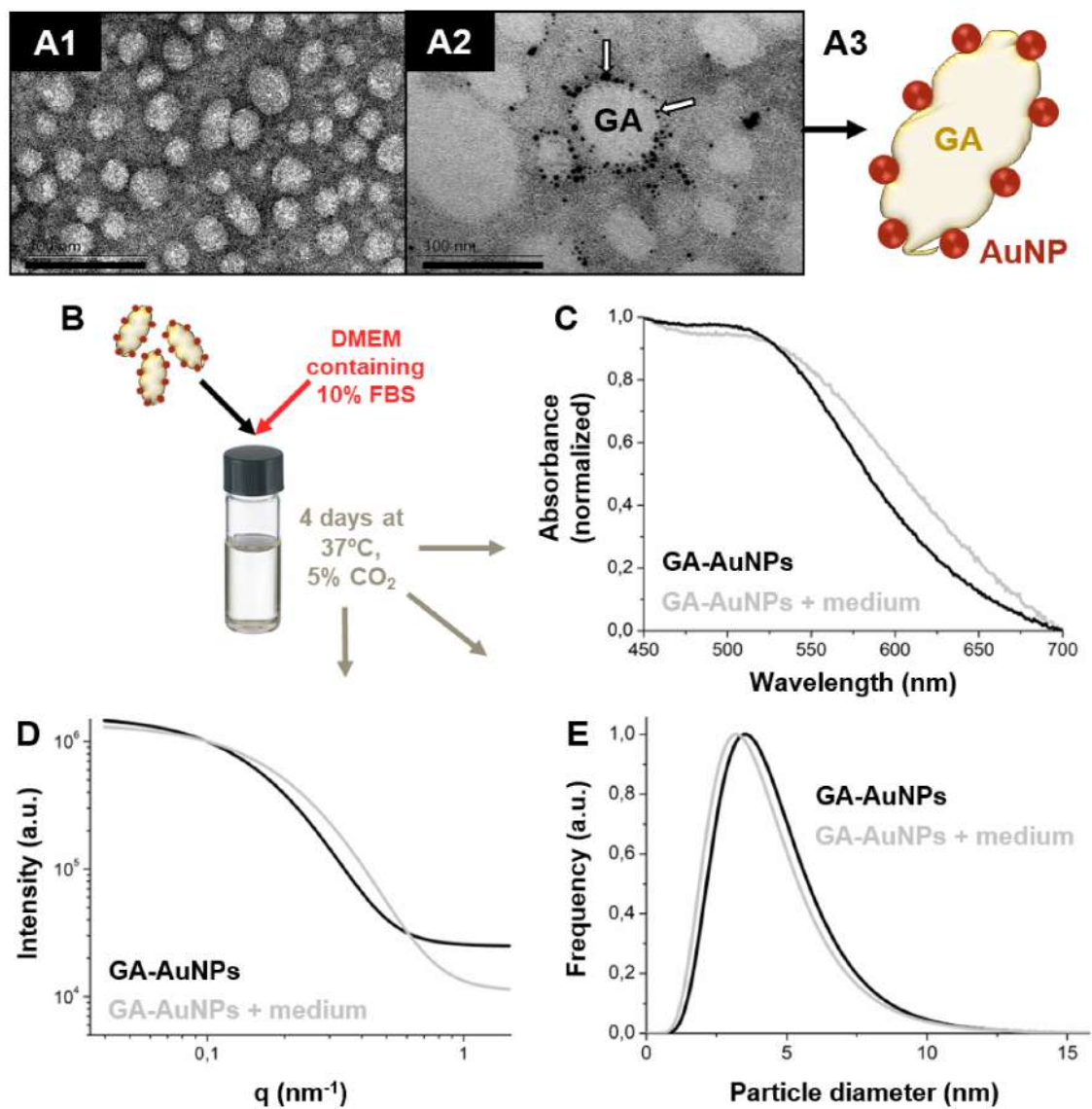

**Figure 1. GA-AuNPs characterization and stability.** Either GA **(A1)** or GA-stabilized AuNPs **(A2)** were negatively stained by uranyl acetate and observed by TEM. Arrows indicate AuNPs. Scale bar = 100 nm. GA-stabilized AuNPs 3D digital representation – out of scale. **(A3)**. GA-AuNPs were mixed with complete cell culture media and kept for 4 days under culture conditions **(B)**. After 4 days, absorbance and X-ray scattering were analyzed by UV-vis spectroscopy **(C)** and SAXS **(D and E)**. Best fits of AuNPs SAXS experimental curves **(D)** and size distributions **(E)** in water and in cell culture medium.

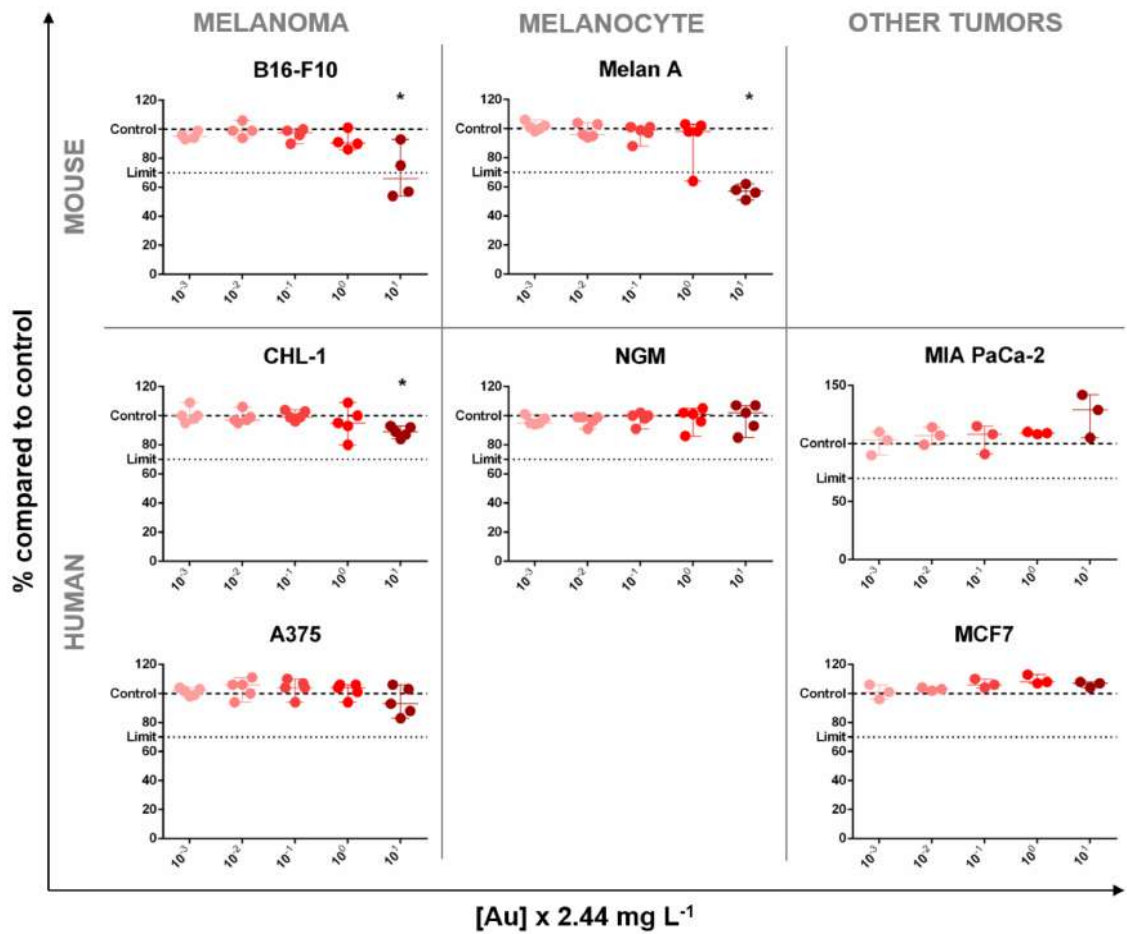

**Figure 2. GA-AuNPs cytotoxicity screening.** Tumor and non-tumor cell lineages were exposed to a range of AuNPs concentrations for 96 hours. Cells were submitted to the MTT assay. Exposed samples were compared to the control (0.4% GA – dashed lines) of each experiment. Dotted lines represent 30% cell viability loss. Data from at least 3 independent experiments are presented as median with range. Paired t-test. *p<0.05.

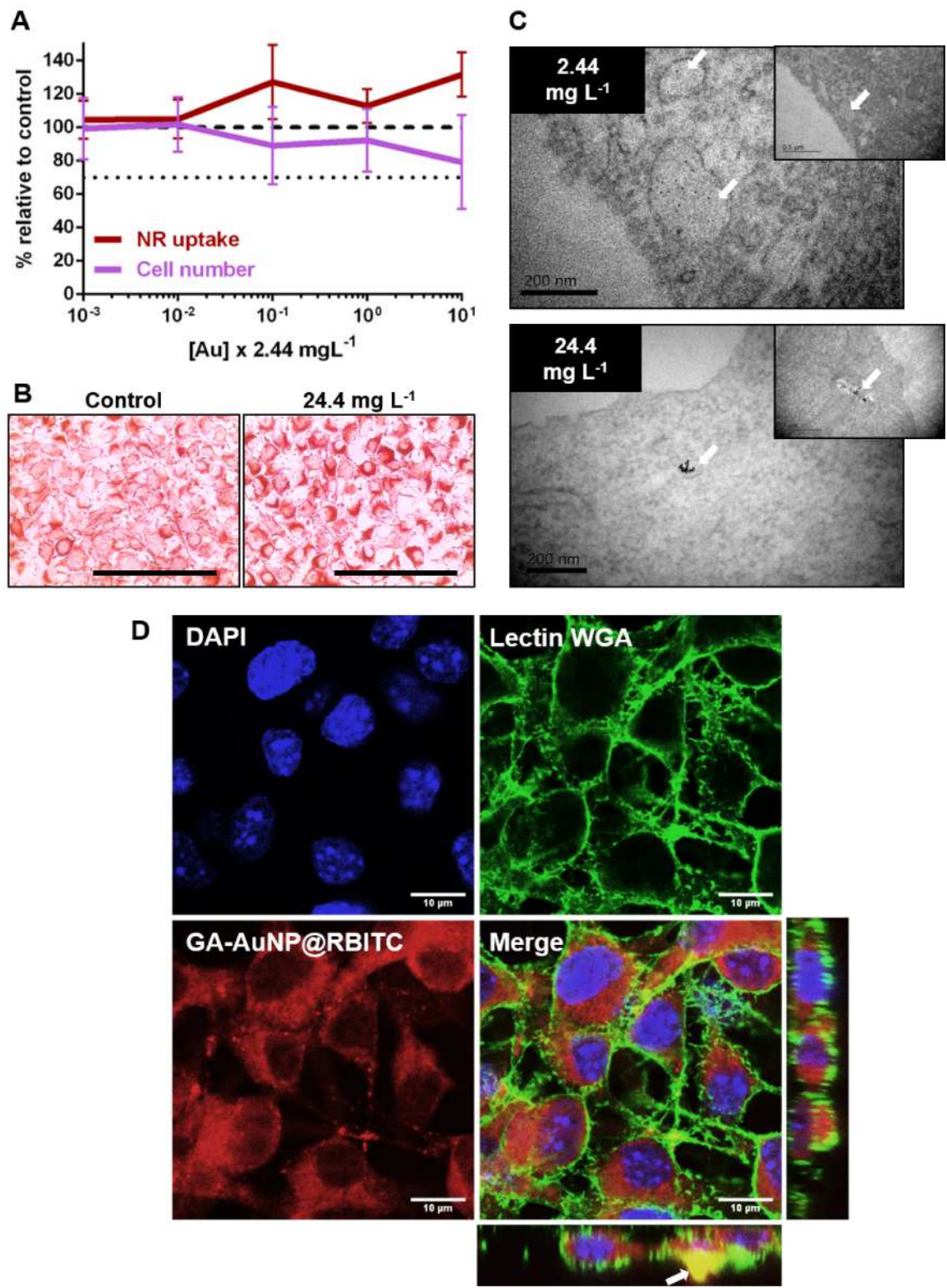

**Figure 3. GA-AuNPs effect on melanoma cells uptake and proliferation.** B16-F10 cells were exposed to a range of GA-AuNPs concentrations for 96 hours. Neutral red (NR) uptake **(A and B)** and cell number **(A)** were colorimetrically determined. NR-absorbance was normalized by each sample cells number. Exposed samples were compared to control (0.4% GA – dashed line in A) of each experiment. Dotted line in A represents 30% loss in any of the parameters. Data from at least 5 independent experiments are presented as mean ± SD. Paired t-test, α=0.05. After NR incubation, cells were observed by phase contrast (scale bar = 160 μm) **(B)** and TEM (arrows = AuNPs inside intracellular vesicles; scale bar = 200 nm) **(C)**. **D)** B16-F10 cells were cultivated in the presence of 2.44 mg L$^{-1}$ GA-AuNP@RBITC, then labeled with DAPI and Lectin WGA 488 and imaged by laser scanning confocal microscopy. Inserts perpendicular to the merged image show the orthogonal view of a specific spot. Arrow = overlap region between Lectin WGA and GA-AuNP@RBITC.

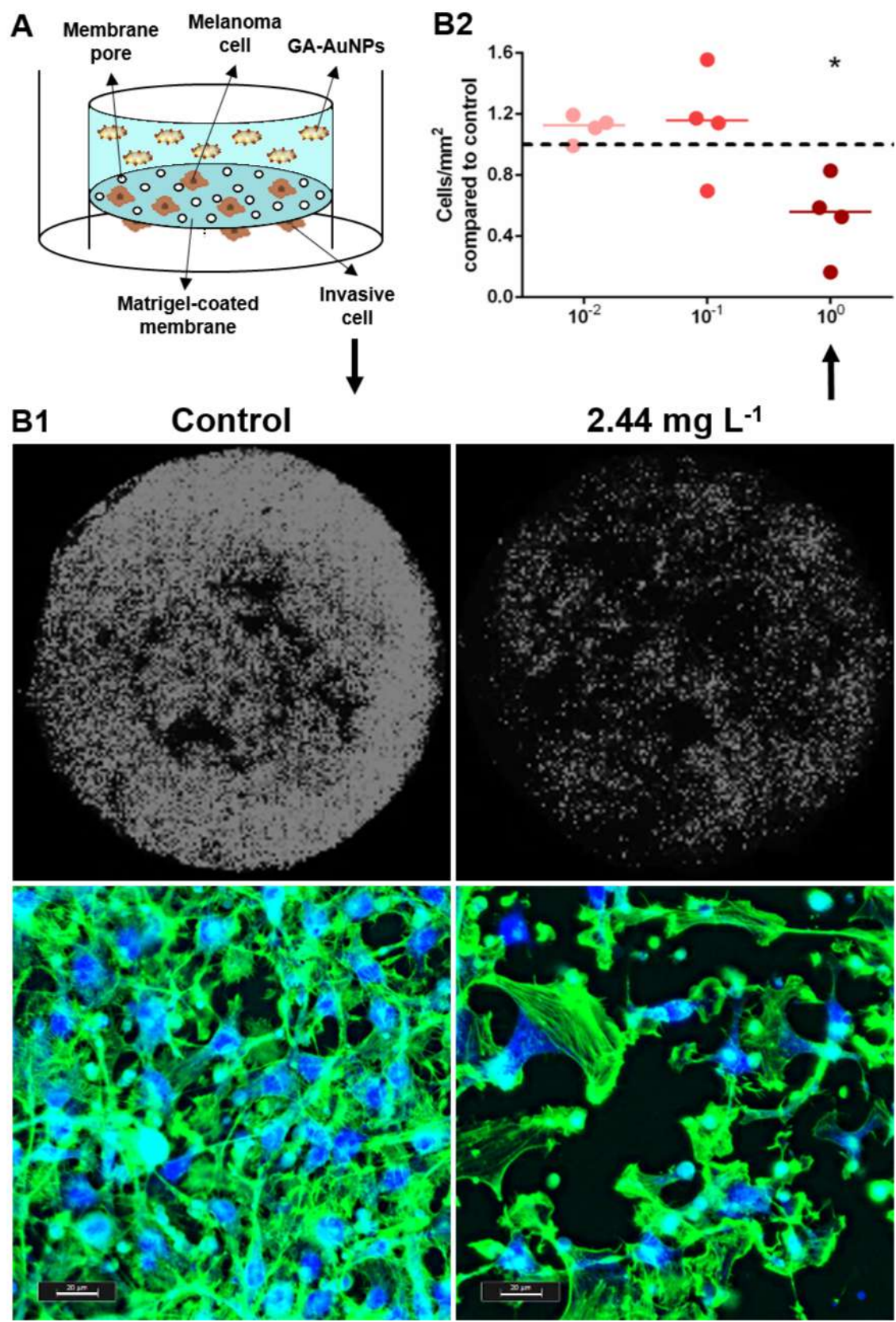

**Figure 4. GA-AuNPs effect on melanoma cells invasive capacity. A)** Experimental

scheme (not drawn to scale). **B1)** On top, representative images showing the total membrane area with invasive cells nuclei stained with DAPI. Bottom images show representative areas with cells actin cytoskeleton stained with ActinGreen 488 (upper images). Scale bar = 20 μm. **B2)** Nuclei number per area was quantified using Fiji software, and data from 4 independent experiments are presented. Exposed samples were compared to control (0.4% GA – dashed line) of each experiment. Graph bars represent data median. Paired t-test. *p<0.05.

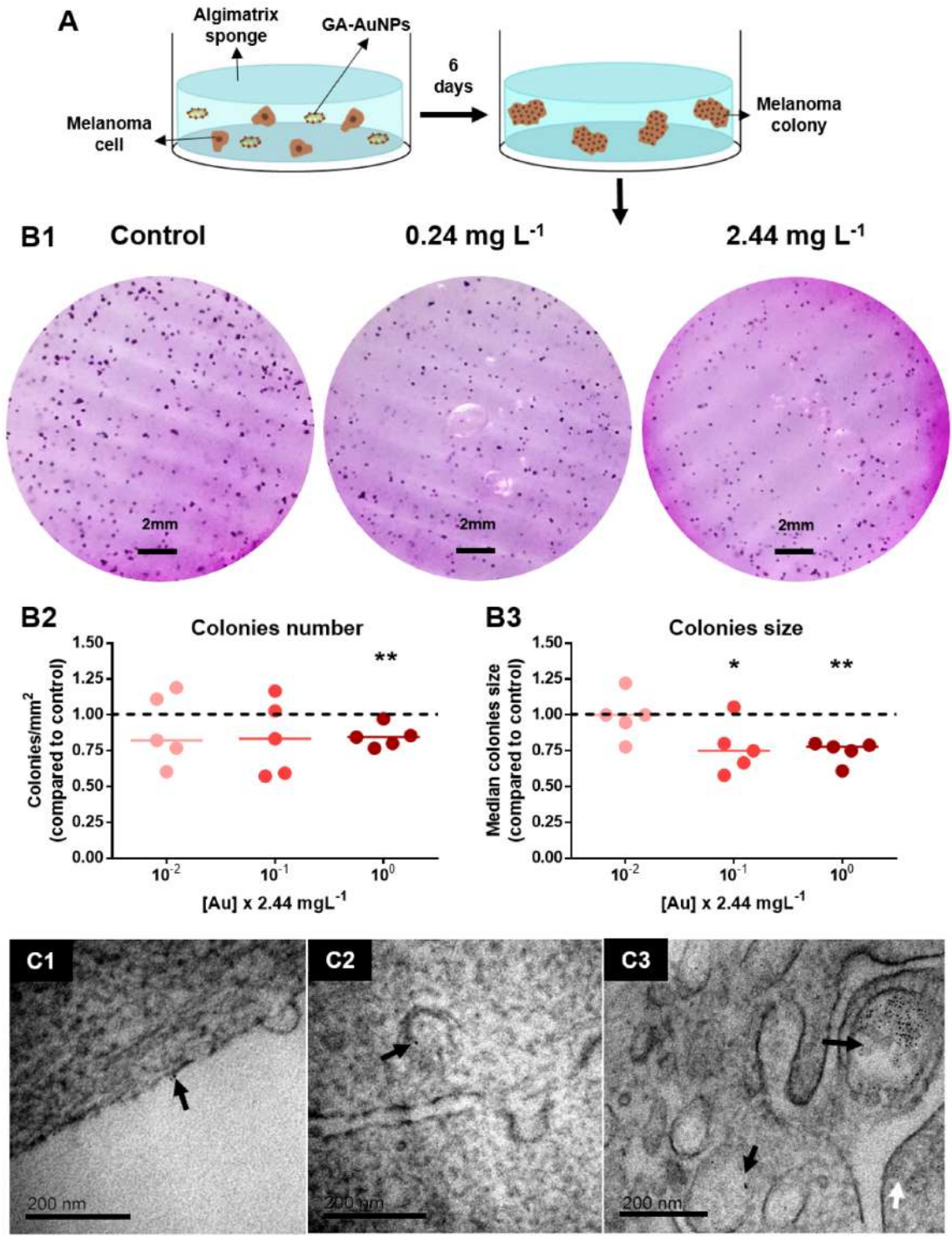

**Figure 5. GA-AuNPs effect on melanoma cells clonogenic capacity.** Experimental scheme (not drawn to scale) **(A).** Representative stained colonies-containing hydrogels **(B1)**. Colonies number **(B2)** and size **(B3)** quantification using Fiji software. Exposed samples were compared to control (0.4% GA – dashed line) of each experiment. Data

from 5 independent experiments. Graph bars represent data median. Paired t-test. *$p<0.05$, **$p<0.01$. Colonies harvested from Algimatrix sponges and processed for TEM **(C 1-4)**. Arrows show AuNPs localized at the cell surface **(C1)**, inside endocytic **(C2)** or cytosolic **(C3, black arrow)** vesicles and free in cytoplasm **(C3, white arrow)**. Scale bars = 200 nm.

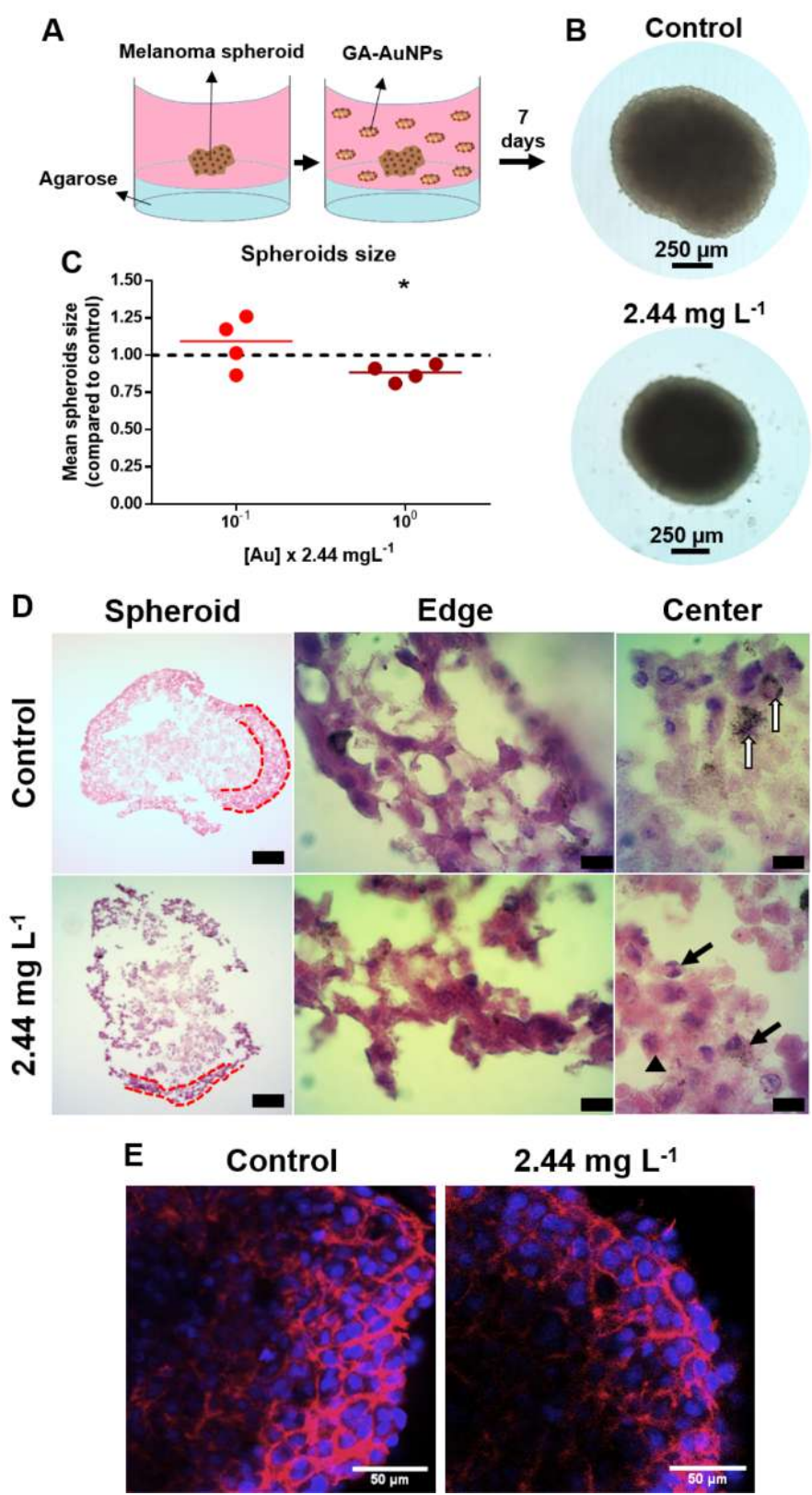

**Figure 6. GA-AuNPs effect on melanoma spheroids.** Experimental scheme (out of scale) **(A).** Representative spheroids **(B)**. Spheroids size quantification **(C)** was performed using Fiji software. Exposed samples were compared to control (0.4% GA – dashed line) of each experiment. Data from 4 independent experiments is presented. Graph bars represent data median. Paired t-test. *p<0.05. **D)** Histologically processed spheroids stained with hematoxylin and eosin. Representative regions under light microscopy. Dashed lines highlight representative spheroids edge thickness. White arrows show melanin producing cells. Black arrows and arrow heads show morphologically altered cells. Scale bars for panoramic images = 100 µm, for detailed images = 10 µm. **E)** Representative region of spheroids edge under confocal fluorescence microscopy showing cells nuclei stained with PI (pseudocolored in blue) and actin cytoskeleton with Alexa Fluor 647 phalloidin (pseudocolored in red).

**TABLE**

**Table 1. Quantification of GA-AuNPs uptake by melanoma cells.**

| [Au] to which cells were exposed (mg L$^{-1}$) | Au mass (ng) uptaken by $10^6$ cells | Number of GA-AuNPs uptaken by each cell |
|:---:|:---:|:---:|
| 2.44 | 3.99 ± 0.82 | 16,257.75 ± 3,334.79 |
| 24.4 | 27.09 ± 6.03 | 110,136.05 ± 24,516.24 |

Result of 3 independent experiments. Data represent mean ± SD. The number of GA-AuNPs uptaken by each cell was calculated based on the mass values of an individual AuNP (2.46 x 10$^{-10}$ ng).

**SUPPORTING INFORMATION**

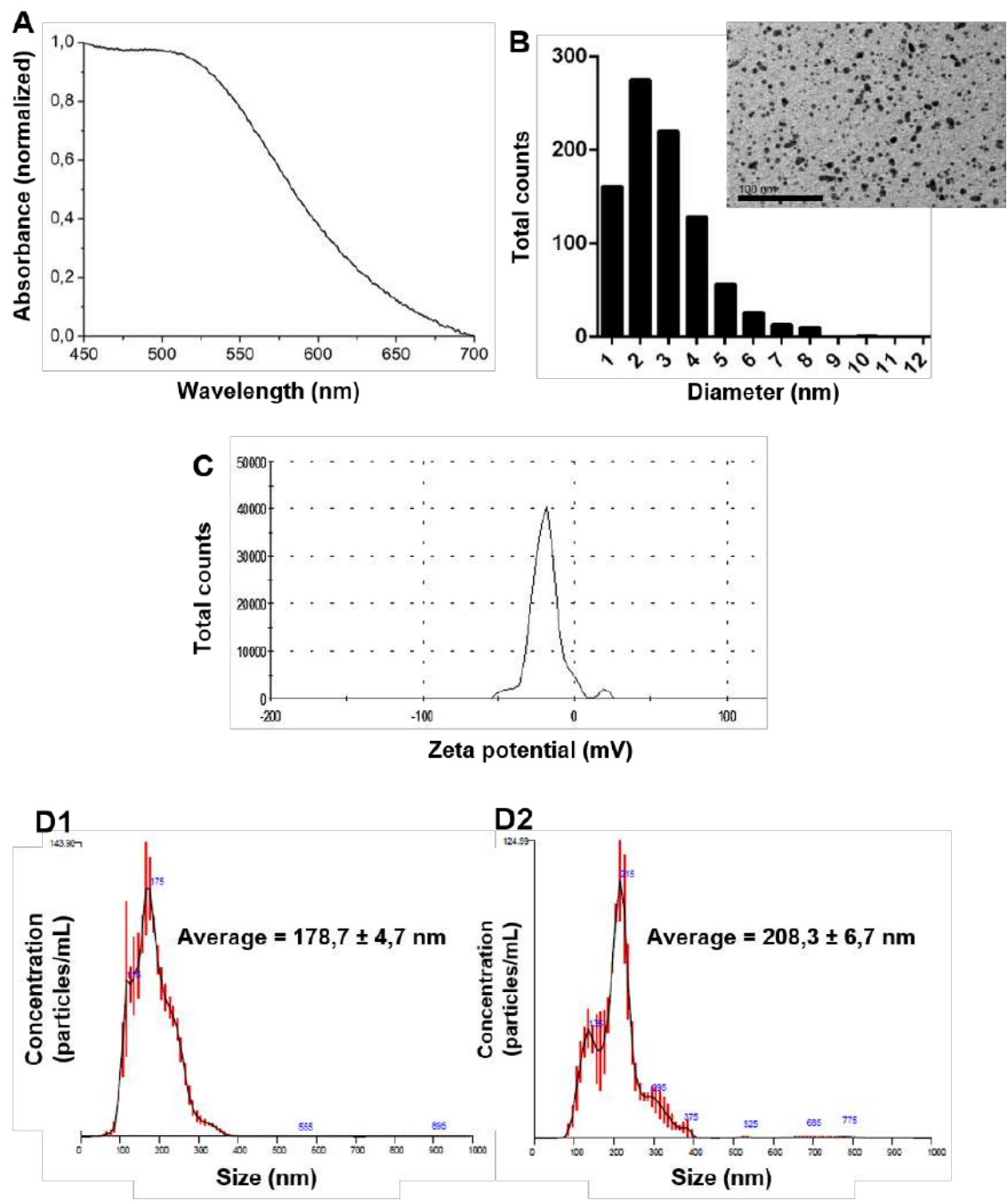

**Supplementary figure 1. GA-AuNPs characterization.** AuNPs were chemically synthesized by HAuCl$_4$ reduction using NaBH$_4$ and simultaneously stabilized by GA. **A)** Normalized UV-vis spectroscopy. **B)** TEM images were obtained (scale bar = 100 nm) and the individual AuNPs diameter was quantified using Fiji software. **C)** Zeta potential

distribution. **D)** Aqueous dispersions of GA (1) or GA-stabilized AuNPs (2) were analyzed on NanoSight equipment for GA hydrodynamic size in solution.